\documentclass[12pt]{article}

\usepackage{cite,amsmath,amsfonts,amsthm,fullpage}

\newcommand{\Pf}{\mathop\mathrm{Pf}\nolimits}
\newcommand{\sgn}{\mathop\mathrm{sgn}\nolimits}
\newcommand{\DP}{\mathop\mathrm{DP}\nolimits}
\def\bt{{\mathbf t}}
\usepackage{amssymb}
\usepackage{amsbsy}
\usepackage{epsfig}
\usepackage{verbatim}
\makeatletter \@addtoreset{figure}{section}
\def\thefigure{\thesection.\@arabic\c@figure}
\def\fps@figure{h, t}
\@addtoreset{table}{bsection}
\def\thetable{\thesection.\@arabic\c@table}
\def\fps@table{h, t}
\@addtoreset{equation}{section}

\newtheorem{corollary}{Corollary}[section]
\newtheorem{definition}{Definition}[section]

\newtheorem{proposition}{Proposition}[section]

\newtheorem{examps}{Examples}[section]

\newtheorem{lemma}{Lemma}[section]
\newtheorem{remark}{Remark}[section]
\newtheorem{remarks}[remark]{Remarks}

\def\bx{\begin{example}}
\def\ex{\end{example}}
\def\bxs{\begin{examps}. \rm\begin{enumerate}}
\def\exs{\end{enumerate}\end{examps}}
\def\bd{\begin{definition}}
\def\ed{\end{definition}}
\def\bp{\begin{proposition}\rm}
\def\ep{\end{proposition}}
\def\bc{\begin{corollary}}
\def\ec{\end{corollary}}
\def\bl{\begin{lemma}\em}
\def\el{\end{lemma}}
\def\be{\begin{equation}}
\def\ee{\end{equation}}
\def\br{\begin{remark}\rm\small}
\def\er{\end{remark}}
\def\brs{\begin{remarks}.\\ \rm\
\begin{enumerate}}
\def\ers{\end{enumerate}\end{remarks}}
\def\bea{\begin{eqnarray}}
\def\eea{\end{eqnarray}}
\def\det{\mathrm {det}}
\def\sgn{\mathrm {sgn}}
\def\ln{\mathrm {ln}}

\def\&{&{\hskip -20pt}}
\def\bs{{\mathbf s}}

\def\p{{\phi}}
\def\l{\langle}
\def\r{\rangle}

\def\Nb{{\mathbf N}}

\date{april 26}

\begin{document}
%\begin{flushright}
%CRM-3313 (2011)
%\end{flushright}

\medskip
\begin{center}
\begin{Large}\fontfamily{cmss}
\fontsize{17pt}{27pt} \selectfont \textbf{Multiple sums and
integrals  as \\ neutral BKP tau functions}\footnote{This work was
 partially supported by the European Union, through
the FP6 Marie Curie RTN ENIGMA (Contract no.MRTN-CT-2004-5652) and
the European Science Foundation Program MISGAM 8,
by the Natural Sciences and Engineering Research
Council of Canada (NSERC), the Fonds FCAR du Qu\'ebec,
 and  by the Russian Academy of Science program ``Fundamental
Methods in Nonlinear Dynamics'', RFBR-Italian grant No
09-01-92437-KE-a and RFBR grant 08-01-00501.}
\end{Large}\\
\bigskip
\begin{large}
{J. Harnad}$^{\dagger\flat}$ \footnote{harnad@crm.umontreal.ca},
 {J.W. van de Leur}$^{\star}$\footnote{J.W.vandeLeur@uu.nl}
 and
 {A. Yu. Orlov}$^{\star\star}$\footnote{ orlovs55@mail.ru}
\end{large}
\\
\bigskip
\begin{small}

 $^{\dagger}$ {\em Centre de recherches
math\'ematiques, Universit\'e de Montr\'eal\\ C.~P.~6128, succ.
centre ville, Montr\'eal,
Qu\'ebec, Canada H3C 3J7} \\
\smallskip
$^{\flat}$ {\em Department of Mathematics and Statistics,
Concordia University\\ 7141 Sherbrooke W., Montr\'eal, Qu\'ebec,
Canada H4B 1R6} \\
\smallskip
$^{\star}${\em Mathematical Institute,University of Utrecht,\\
P.O. Box 80010, 3508 TA Utrecht,
The Netherlands}\\
\smallskip
$^{\star\star}$ {\em Nonlinear Wave Processes Laboratory, \\
Oceanology Institute, 36 Nakhimovskii Prospect\\
Moscow 117851, Russia } \\
\end{small}
\end{center}
\bigskip
\bigskip

%%%%%%%%%%%%%%%% Abstract %%%%%%%%%%%%%%%%
\begin{center}{\bf Abstract}
\end{center}
\smallskip

\begin{small} We consider multiple sums and multi-integrals as
tau functions of the BKP hierarchy using neutral fermions as the
simplest  tool for deriving these. The sums are over projective
Schur functions $Q_\alpha$ for strict partitions $\alpha$. We
consider two types of such sums: weighted sums of $Q_\alpha$ over
strict partitions $\alpha$ and sums over products $Q_\alpha
Q_\gamma$ . In this way we obtain discrete analogues of the
beta-ensembles ($\beta=1,2,4$). Continuous versions are
represented as multiple integrals. Such sums and integrals are of
interest in a number of problems in mathematics and physics.

\end{small}

\bigskip

%%%%%%%%%%%%%%%% Section 1  Introduction %%%%%%%%%%%%%%%%

\section{Introduction}

This work deals with certain multiple sums ((\ref{S0})-(\ref{S5}), (\ref{S1-DI})-(\ref{S5-DI}))
and multiple integrals ((\ref{I1})-(\ref{I4}), (\ref{I5})) that
are of interest in a number of problems in mathematics and
physics. The multiple sums appear in models of random
partitions (as developed by A.Vershik's school; see
\cite{Ok-rand-part} for a review), and random motion of particles
\cite{F} (see also \cite{Forr} for a review). The multiple integrals
yields certain analogues of $\beta=1,2,4$ ensembles of random matrices,
 and also an analogue of the two-matrix models \cite{Mehta}. We
consider deformations of the measure defined in terms of four
semi-infinite sets of parameters and relate the generating
function $Z$ to the coupled two-component BKP hierarchy.  The main
tool is the use of $1$- and $2$-component neutral fermions, which
provide links with integrable systems known as neutral
 BKP hierarchies  \cite{DJKM'},\cite{JM}.
One version of these was first introduced  in
\cite{DJKM'},\cite{JM}. In this work, we use another version
introduced in \cite{KvdLbispec}.

%%%%%% Section 2 Sums over projective Shur functions  %%%%%%%%%%%

\section{Sums over projective Schur functions}

In the following, we consider sums over strict partitions , which will
 be denoted by Greek letters $\alpha$, $\beta$. Recall  \cite{Mac} that a
strict partition $\alpha$ is a set of integers (parts)
$(\alpha_1,\dots,\alpha_k)$ with $\alpha_1>\dots
>\alpha_k\ge 0$. The length of a partition $\alpha$, denoted
$\ell(\alpha)$, is the number of nonvanishing parts, thus it is
either $k$ or $k-1$. Let DP be the set of strict partitions
(i.e., with distinct parts).
We also need a subset of  DP, which will be denoted  $DP^2$  and consists of all partitions of the form
$(\alpha_1,\alpha_1-1,\alpha_3,\alpha_3-1,\dots,\alpha_{k-1},\alpha_{k-1} -1)$.
 Consider the following sums (for $L\in \Nb^+$, $\bt := (t_1, t_3, \dots )$, $\bt^*:=(t_1^*, t_3^*, \dots )$,  $\bar{\bt} := (\bar{t}_1, \bar{t}_3, \dots )$).
 \bea
 \label{S0}
S_0(\bt,L)&:=&\sum_{\alpha\in\DP \atop \alpha_1\le
L}\,Q_\alpha(\tfrac 12\bt)
\\
 \label{S1}
S_1(\bt,\bt^*)&:=&\sum_{\alpha\in\DP}\,e^{-U_\alpha(\bt^*)}Q_\alpha(\tfrac
12\bt)
 \\
 \label{S2}
S_2({\bt},{\bar\bt},\bt^*)&:=&\sum_{\alpha\in\DP}\,e^{-U_\alpha(\bt^*)}Q_\alpha(\tfrac
12\bt)Q_\alpha(\tfrac 12{\bar\bt})
 \\
 \label{S00}
S_{00}({\bt},{\bar\bt},L) &:=&\sum_{\alpha\in\DP \atop
\alpha_1\le L}\,Q_\alpha(\tfrac 12\bt)Q_\alpha(\tfrac 12{\bar\bt})
\\
 \label{S3}
S_3(\bt,{A}^c)&:=&\sum_{\alpha\in\DP}\,{A}^c_\alpha
\,Q_\alpha(\tfrac 12\bt)
\\
 \label{S4}
S_{4}(\bt,\bt^*)&:=&\sum_{\alpha\in\DP^2}\,e^{-U_\alpha(\bt^*)}Q_\alpha(\tfrac
12\bt)
\\
 \label{S5}
S_5({\bt},{\bar\bt},D)&:=&\sum_{\alpha,\beta\in\DP \atop
\ell(\alpha)=\ell(\beta)}\,Q_\alpha(\tfrac
12\bt)D_{\alpha,\beta}Q_\beta(\tfrac 12{\bar\bt})
 \eea

 Here, the {\em  projective Schur functions} $Q_\alpha$  are weighted polynomials in the variables
 $\bt=(t_1,t_3,t_5,\dots)$, $\deg t_m =m$, labeled by strict partitions
(See \cite{Mac} for their detailed definition.) Each
 $Q_\alpha(\tfrac 12\bt)$ is known to be a BKP tau function
 \cite{DJKM'},\cite{JM}. (This was a nice observation of
 \cite{You},\cite{Nimmo}).  The fact that only
 odd subscripts appear in the BKP higher times $t_{2m-1}$ is
 related to the reduction from the KP hierarchy.
The coefficients $U_\alpha$ are defined as
\be
 U_\alpha := \sum_{i=1}^k U_{\alpha_i},
 \ee
  where
  \be
  \label{Un}
U_n\,:=\,U_n^{(0)}-\sum_{m\neq 0,{\rm{odd}}} n^m t^*_m - \ln\,n!,\quad n\in \Nb^+
  \ee
  for some given set of constants $\{U_n^{(0)}\}$.

   The coefficients $A^c_\alpha$  on the  right hand side of (\ref{S3}) are
   determined in terms a pair $(A, a)$ where $A$ is an infinite skew symmetric matrix and $a$
an infinite vector.  For a strict partition $\alpha=
(\alpha_1,\dots,\alpha_k )$, where $\alpha_k>0$, the numbers
$A^c_\alpha$ are defined as the Pfaffian of an antisymmetric $2n
\times 2n$ matrix ${\tilde A}$  as follows:
  \be
  \label{A-c}
A^c_{\alpha}:=\,\Pf[{\tilde A}]
  \ee
where for $k=2n$ even
  \be
  \label{A-alpha-even-n}
{\tilde A}_{ij}=-{\tilde A}_{ji}:=A_{\alpha_i,\alpha_j},\quad 1\le
i<j \le 2n
  \ee
and for $k=2n-1$ odd
\be
\label{A-alpha-odd-n} {\tilde A}_{ij}=-{\tilde A}_{ji}:=
\begin{cases}
A_{\alpha_i,\alpha_j} &\mbox{ if }\quad 1\le i<j \le 2n-1 \\
a_{\alpha_i} &\mbox{ if }\quad 1\le i < j=2n  .
 \end{cases}
  \ee
In addition we set $A_0^c =1$.

The coefficients $D_{\alpha,\beta}$ in (\ref{S5}) are defined as
determinants:
 \be\label{D-alpha-beta}
D_{\alpha,\beta}\,=\,\det\,\left( D_{\alpha_i,\beta_j}\right)
 \ee
 where $D$ is a given constant infinite matrix.

\br\label{specification} Series (\ref{S0}), (\ref{S1}), (\ref{S2})
and (\ref{S00}) may be obtained via specializations of $A^c$ in
the series (\ref{S3}).  If we put
\be
A_{nm}=\frac 12 e^{-U_m-U_n}
\sgn(n-m), \qquad a_n=e^{-U_n}
\ee
we obtain (\ref{S1}). If we further choose $U_n=0$ and $U_n =+\infty$  for $n\le L$, $ n>L$
respectively, we obtain (\ref{S0}). If we set
 \be
 A_{nm}=\frac 12
e^{-U_m-U_n}Q_{(n,m)}(\tfrac 12{\bar\bt}), \qquad
a_n=e^{-U_n}Q_{(n)}(\tfrac 12{\bar\bt})
 \ee
 we obtain (\ref{S2}).
Choosing again $U_n = 0$, $U_n = +\infty$ for $n\le L ,n>L$, we
obtain (\ref{S00}). The series (\ref{S4}) is obtained from
(\ref{S3}) by taking
 \be
 A_{nm}=\delta_{n+1,m}-\delta_{m+1,n}.
 \ee
The sums (\ref{S00}) and (\ref{S2}) may be also obtained as
particular cases of (\ref{S5}) by putting
 \be
D_{nm}=e^{-U_m-U_n}\delta_{n,m}
 \ee
to get (\ref{S2}).
\er

 All these sums are particular examples of  BKP tau  functions,
 as introduced in \cite{DJKM'}, defining solutions to what was called the neutral BKP
 hierarchy in \cite{KvdLbispec}.
They may be further specialized if
 we choose $\bt=\bt_\infty :=(1,0,0,\dots)$. Then
  \be
Q_\alpha(\tfrac
12\bt_\infty)=\Delta^*(\alpha)\,\prod_{i=1}^k\,\frac{1}{\alpha_i!}\,,
\quad \alpha=(\alpha_1,\dots,\alpha_k)
  \ee
 where
  \be
 \Delta^*(\alpha)=\Delta^*_k(\alpha):=
 \prod_{0<i<j\le k}\,\frac{\alpha_i-\alpha_j}{\alpha_i+\alpha_j}
 \ee
  For this specialization, we have
 \bea
 \label{S1-DI}
S_1(\bt_\infty,\bt^*) &=& \sum_{\alpha\in\DP}\,\Delta^*(\alpha)
\,\prod_{i=1}^k\,\frac{e^{-U_{\alpha_i}(\bt^*)}}{\alpha_i!}
\\
 \label{S2-DI}
S_2({\bt_\infty},{\bt}_\infty,\bt^*)&=&\sum_{\alpha\in\DP}\,\Delta^*(\alpha)^2
\,\prod_{i=1}^k\,\frac{e^{-U_{\alpha_i}(\bt^*)}}{(\alpha_i!)^2}
\\
 \label{S4-DI}
S_{4}({\bt}_\infty,\bt^*) &=&
\sum_{\alpha\in\DP'}\,{\tilde{\Delta}}^*(\alpha)^4
\,\prod_{i=1}^k\,\frac{e^{-U_{\alpha_i}(\bt^*)-U_{\alpha_i+1}(\bt^*)}}
{\alpha_i!(\alpha_i+1)!}
\\
 \label{S5-DI}
S_5({\bt_\infty},{\bt}_\infty,\bt^*) &=& \sum_{k=0}^\infty \,\,
\frac {1}{k!}\sum_{\alpha,\beta\in\DP\atop
\ell(\alpha)=\ell(\beta)=k}\,
\Delta^*(\alpha)\Delta^*(\beta)\,\prod_{i=1}^k \, \frac
{D_{\alpha_i,\beta_i}}{\alpha_i!\beta_i!}
 \eea

Here $DP'$ is the set of all strict partitions
$(\alpha_1,\alpha_2,\dots,\alpha_N>0)$ with the property
$\alpha_{i}>\alpha_{i+1} +1,\,i=1,\dots,N-1$, and
 \be
{\tilde{\Delta}}^*(\alpha)^4\,:=\,\prod_{i<j\le N}
\frac{(\alpha_i-\alpha_j)^2\left((\alpha_i-\alpha_j)^2-1\right)}
{(\alpha_i+\alpha_j)^2\left((\alpha_i+\alpha_j)^2-1\right)} .
 \ee

If we replace the term $\Delta^*$ in the above  by the Vandermonde
determinant $\Delta$,  the resulting sums
(\ref{S1-DI})-(\ref{S5-DI}) may be viewed as discrete analogues of
matrix models (see \cite{OS}).

\paragraph{Applications.}
The following are some examples of applications of the above sums.

\medskip
\noindent (I) The sum (\ref{S00}) was considered first by Tracy
and Widom in \cite{TW} in a study of the shifted Schur measure.

\medskip
\noindent (II) Sums (\ref{S1}) and (\ref{S2}) may be viewed as
generalizations of hypergeometric functions for the case of many
variables. For example, a generalization of the hypergeometric
function of type ${_pF}_r$ may be obtained from expressions
(\ref{S1}) and (\ref{S2}) by defining the parameters $U_n$ in
terms of Gamma functions as follows:
 \be
 U_n\,=\,\log\,\frac{\prod_{i=1}^p\Gamma(n+a_i)}{\prod_{i=1}^r\Gamma(n + b_i)}
  \ee
 for some set of $p+r$ constants $\{a_i, b_i\}$.
 Sums of type (\ref{S2}) were considered in \cite{Q}, while sums
 (\ref{S1}) are new. (This case will be considered in detail elsewhere \cite{sQQ}.)
 It may be shown that both series (\ref{S1})
 and (\ref{S2}) may be expressed as Pfaffians of matrices whose entries are
 expressed via ${_pF}_r$ and share many properties with the usual
 hypergeometric functions ${_pF}_r$.  Analogues of basic hypergeometric
  functions may be obtained in a similar way.

  \medskip
\noindent (III) Consider models of random strict partitions $\alpha$,
 where the relative weight $W_\alpha$ is given by one of
the following:

\smallskip \noindent(A)
\be W_\alpha\,= \,{A}^c_\alpha \,Q_\alpha(\tfrac 12\bt), \ee
 where $A^c=(A,a)$ and $\bt=(t_1,t_3,\dots)$ are parameters of the model

\smallskip \noindent (B)
\be W_\alpha\,=\,e^{-U_\alpha(\bt^*)}Q_\alpha(\tfrac 12\bt), \ee
where the parameters are $\bt=(t_1,t_3,\dots)$
and $U=(U_0,U_1,\dots)$.

\smallskip \noindent (C)
\be W_\alpha\,=\,e^{-U_\alpha(\bt^*)}Q_\alpha(\tfrac
12\bt)Q_\alpha(\tfrac 12{\bar \bt}), \ee where
$\,\bt=(t_1,t_3,\dots)$, ${\bar\bt}=({\bar t}_1,{\bar t}_3,\dots)$
and $U=(U_0,U_1,\dots)$ are independent parameters. (Note that models (B) and
(C) are particular cases of model (A).)

The series $S_3$, $S_1$ and $S_2$ may then be be viewed as
normalization factors (partition functions) respectively for
models (A),(B), and (C). Similarly,  series $S_5$ is a partition
function for a model of strict bi-partitions.

  \medskip \noindent (IV) Series $S_1$ and $S_2$ were used in \cite{LO}, where
random oscillating Young diagrams related to strict partitions were considered.

 \br
 The fermionic representation for these models allows their correlation functions
to be computed  in  standard ways. (See e.g. refs.
\cite{Okoun-cor},\cite{Foda},\cite{Serbian} and
\cite{HO-ration-ferm}.)
 \er

\paragraph{Neutral fermions.}
To construct tau functions of the BKP and  two-component BKP
hierarchies (see \cite{JM},\cite{KvdLbispec}) we need
 \emph{neutral free fermions} $\{ \phi_i^{(1)}$, $\phi_i^{(2)}\}_{i\in {\bf
Z} }$, satisfying the anticommutation relations
\begin{equation}\label{canonical}
    [\phi_n^{(a)},\phi_m^{(b)}]_+\,=\,(-1)^n\delta_{n,-m}\delta_{a,b}\,,\quad
    a,b=1,2.
\end{equation}
For one-component BKP only the first component
$\p_n:=\p_n^{(1)},\, n\in\mathbb{Z}$ is used. The fermionic Fock
space will be chosen as in \cite{KvdLbispec}) (as opposed to
\cite{JM}). Namely, the action of neutral fermions on vacuum states
is defined by
 \bea
 \label{phi-on-vacuum}
\phi_n^{(a)}|0\rangle &=&0, \qquad   \qquad  \ \langle
0|\phi_{-n}^{(a)}=0,\qquad n<0,
   \\
 \label{zero-phi-on-vacuum} \phi_0^{(a)}|0\rangle&=&\frac{1}{\sqrt
2}|0\rangle, \qquad \langle 0|\phi_0^{(a)}=\frac{1}{\sqrt
2}\langle 0|
 \eea
where relation (\ref{zero-phi-on-vacuum}) is chosen as in
\cite{KvdLbispec}). (See also the Appendix A in the arXiv version
of \cite{LO} for some details on links between \cite{JM} and
\cite{KvdLbispec}.)

For linear combinations
\be w_k=\sum_a\sum_{n}c^{(a)}_{k,n}\p^{(a)}_n, \quad k-1, 2, \dots ,
\ee
Wick's Theorem implies, for arbitrary such products of an even number of $w_k$'s
 \be
 \langle w_1
\cdots w_{2n} \rangle =\sum_{\sigma \in S_{2n} } \mbox{sign}(\sigma) \langle
w_{\sigma(1)}w_{\sigma(2)}\rangle \cdots \langle
w_{\sigma(2n-1)}w_{\sigma(2n)} \rangle
 =:\rm{Pf}\left(\left(\langle w_iw_j\rangle\right)_{1\le i,j\le 2n}\right)\, .
  \ee
where the sum is over the permutation group $S_{2n}$.

We also need the Fermi fields
\begin{equation}\label{phifields}
\phi^{(a)}(z)=\sum_{n\in\mathbb{Z}}\phi^{(a)}_nz^n,\qquad a=1,2
\end{equation}
To evaluate integrals we use
 \be
 \label{phi-phi}
\langle 0|\phi^{(b)}(z_1)\phi^{(a)}(z_2)|0
\rangle={\frac{1}{2}}\left(\frac{z_1-z_2}{z_1+z_2} \right)
\delta_{ab}.
 \ee
For $m\in \Nb^+$ Wick's theorem implies
 \be
 \label{FermiDelta}
\l 0| \phi^{(a)}(z_1)\phi^{(a)}(z_2)\cdots\phi^{(a)}(z_{m}) |0\r
=\left(\frac12\right)^{\frac m2}\Delta^*_{m}(z)
 \ee
 Note that, because of (\ref{zero-phi-on-vacuum}), this expectation value is
 nonvanishing for $m$ odd.

\paragraph{BKP tau functions.}

The general tau function of the two-component 2-BKP hierarchy is expressed
in fermionic form as
 \bea
 \label{2c-2-nBKP}
 &\tau^{2c-2-BKP}(\bt^{(1)},{\bt}^{(2)},{\bar\bt}^{(1)},{\bar\bt}^{(2)},A) \cr
&=
\l0|\,\Gamma^{(1)}(\bt^{(1)})\Gamma^{(2)}(\bt^{(2)})\,e^{\sum_{a,b}\sum_{n,m
}\,A_{nm}^{a,b}\phi_n^{(a)}\phi_m^{(b)}
}\,{\bar\Gamma}^{(1)}({\bar\bt}^{(1)}){\bar\Gamma}^{(2)}({\bar\bt}^{(2)})\,|0\r
.
 \eea
 where
 \be
 \Gamma^{(a)}(\bt^{(a)}):=\exp\, \sum_{n\ge 1,\;\text{odd}}B^a_n t_n^{(a)}\,
 ,\quad {\bar\Gamma}^{(a)}({\bar \bt}^{(a)}):
 =\exp\, \sum_{n\ge
 1,\;\text{odd}}B_{-n}^{(a)}{\bar t}_n^{(a)}
 \ee
 \be
    B_n^{(a)}\, :=\,\frac12\sum_{i\in\mathbb{Z}}(-1)^{i+1}\phi_i^{(a)}\phi_{-i-n}^{(a)}.
 \ee

 The array of numbers $A=\{A_{nm}^{a,b}, \ a,b=1,2;\ n,m\in\mathbb{Z}\}$  is the data
that determine the two-component 2-BKP tau function.
 The four sets of independent parameters,
 ${\bt}^{(a)}=({ t}_1^{(a)},{ t}_3^{(a)},...)$,
 ${\bar\bt}^{(a)}=({\bar t}_1^{(a)},{\bar t}_3^{(a)},...)$, for $a=1,2$, are
 called higher times of the hierarchy. If we fix any three sets,
 the fourth will be the higher times of the usual
 BKP hierarchy \cite{KvdLbispec}. In this sense (\ref{2c-2-nBKP})
 may be viewed as a four coupled BKP tau function. If we set
 ${\bar\bt}^{(1)}={\bar\bt}^{(2)}=0$ we obtain the two-component
 BKP hierarchy, as  described in \cite{KvdLbispec}.

 For the one-component case we omit the second component
and the superscripts. The 2-BKP tau function is then
 \be\label{2-nBKP}
 \tau^{2BKP}(\bt,{\bar\bt})\,=\,\l
0|\,\Gamma(\bt)\,e^{\sum_{n,m \in\mathbb{Z} }\,A_{nm}\phi_n\phi_m
}\,\Gamma({\bar\bt})\,|0\r
 \ee
and the usual neutral BKP tau function may be written
 \be\label{nBKPgeneral}
 \tau^{BKP}(\bt)\,=\,\l
0|\,\Gamma(\bt)\,e^{\sum_{n,m \in\mathbb{Z} }\,A_{nm}\phi_n\phi_m
}\,|0\r
 \ee
A remarkable example of a BKP tau function was found in
\cite{You}; namely, Schur's $Q$-functions $Q_\alpha$ themselves \cite{Mac}. These may be
expressed fermionically as
 \be
 \label{You}
 \langle 0|\,\Gamma(\bt)\,\phi_{\alpha_1}\phi_{\alpha_2}\cdots\phi_{\alpha_{2N}}|0\rangle=
2^{-\tfrac 12\ell(\alpha)}Q_{\alpha}(\tfrac 12\bt)
 \ee
 where $\alpha_1>\alpha_2>\cdots >\alpha_{2N}\ge 0$. The set $(\alpha_1,\dots,\alpha_{2N})$ and the
  partition $\alpha$ are related as follows: in case
  $\alpha_{2N}>0$, $\alpha=(\alpha_1,\dots,\alpha_{2N})$ and $\ell(\alpha)=2N$, while in
  case $\alpha_{2N}=0$, $\alpha=(\alpha_1,\dots,\alpha_{2N-1})$ and $\ell(\alpha)=2N-1$.

\paragraph{Fermionic representation for sums.} The formulae below
show that sums (\ref{S0})-(\ref{S4}) are BKP tau functions. First, we have
 \be
 \label{S3F}
S_3(\bt,{A}^c)\,=\,\sum_{\alpha\in\DP}\,{A}^c_\alpha
\,Q_\alpha(\tfrac 12\bt)\,=\,\l 0|\,\Gamma(\bt)\,e^{2\sum_{n>m
> 0}\,A_{nm}\phi_n\phi_m\,
+\,2\sum_{n>0}\,a_n\phi_n\phi_0}\,|0\r .
 \ee
In view of Remark \ref{specification} we have fermionic
representations for (\ref{S0}), (\ref{S1}), (\ref{S2}), (\ref{S00}).
In particular
 \bea
 \label{S0F} S_0(\bt,N)&\&=\,\sum_{\alpha\in\DP
\atop \alpha_1\le N}\,Q_\alpha(\tfrac 12\bt)=\l
0|\,\Gamma(\bt)\,e^{2\sum_{N\ge n>m
\ge 0}\phi_n\phi_m}\,|0\r
\\
 \label{S1F}
S_1(\bt,\bt^*) &\&=\sum_{\alpha\in\DP}\,e^{-U_\alpha(\bt^*)}Q_\alpha(\tfrac
12\bt) =\l 0|\,\Gamma(\bt)\,\mathbb{T}(\bt^*)\,e^{2\sum_{n>m\ge 0}\phi_n\phi_m}\,|0\r
\\
 \label{S4F}
S_{4}(\bt,\bt^*) &\&= \sum_{\alpha\in\DP^2}\,e^{-U_\alpha(\bt^*)}Q_\alpha(\tfrac
12\bt) =\l 0|\,\Gamma(\bt)\,\mathbb{T}(\bt^*)\,e^{2\sum_{ n>
0}\,\phi_n\phi_{n+1}}\,|0\r
 \eea
where
 \be
\mathbb{T}(\bt^*):=\,\exp\,\left(-\sum_{n>0}\,U_n(\bt^*)\phi_n\phi_{-n}\right)
 \ee

Evaluating at $\bt= \bt_\infty$ gives a ``solitonic'' representation for
 $S_1(\bt=\bt_\infty,\bt^*)$ :
  \be
S_1(\bt_\infty,\bt^*)\,=\,\frac 1c\l
0|\,\Gamma(\bt^*_+)\,e^{2\sum_{n>m
>0 }\,e^{-U_m^{(0)}-U_n^{(0)}}{\phi(n)\phi(m)} + 2\sum_{n>0}
e^{-U_n^{(0)}}{\phi(n)\phi_0} }\,\Gamma({\bt}^*_-)\,|0\r
 \ee
  where Fermi fields (\ref{phifields}) are used (see \cite{LO}).
 This shows that this series is a 2-BKP tau function with respect
 to the higher times $\bt^*_\pm$ defined as $\bt^*_+:=(t_1^*,t_3^*,\dots)$ and
 $\bt^*_-:=(-t_{-1}^*,-t_{-3}^*,\dots)$. Here $c$ is a normalization factor given by
  \be
  \l 0|\Gamma(\bt^*_+) \Gamma({\bt}^*_-)|0\r=b(\bt^*_-,\bt^*_+).
  \ee

\br
Another fermionic representation for (\ref{S2}),(\ref{S00})
was obtained in \cite{Q}.
 \er
 Finally, we have
 \be\label{S5F}
S_5({\bt}^{(1)},{\bt}^{(2)},D)\,=\,
 \l
0|\,\Gamma^{(1)}(\bt^{(1)})\Gamma^{(2)}(\bt^{(2)})\,e^{2\sum_{n,m
> 0}\,D_{nm}\phi_n^{(1)}\phi_m^{(2)} }\,|0\r .
 \ee

%%%%%%%%%%%%%%%% Section 3  Multiple integrals %%%%%%%%%%%%%%%%

\section{Multiple integrals}

Let $d\nu$ be a measure supported on a contour $\gamma$ on the
complex plane. We take $\gamma$ as either of the following two contours:

 (A) An interval on the real axes $0\le z < \infty$

 (B) A segment of the unit  circle: given  by
  $z=e^{i\varphi},\, 0\le \varphi \le\theta$, \ $0<\theta<\pi$,\hfill  \break

  \noindent
Consider the following $N$-fold integrals:
 \be\label{I1}
I_1(N):=\int_\gamma\cdots\int_\gamma\,|\Delta^*(z)|
\,\prod_{i=1}^N\, d\nu(z_i)
 \ee
  \be\label{I2}
   I_2(N):=\int_\gamma\cdots\int_\gamma\,|\Delta^*(z)|^2
\,\prod_{i=1}^N\, d\nu(z_i)
 \ee
 \be\label{I3}
I_3(N):=\int_\gamma\cdots\int_\gamma\,\Delta^*(z)\,a^c({\bf z}) \,
\prod_{i=1}^N\, d\nu(z_i)
 \ee
  \be\label{I4}
   I_4(N):=\int_\gamma\cdots\int_\gamma\,|\Delta^*(z)|^4
\,\prod_{i=1}^N\, d\nu(z_i)
 \ee
where, as before,
\[
\Delta^*(z)=\prod_{i>j}^N\frac{z_i-z_j}{z_i+z_j}
\]
The notation $a^c({\bf z})$ is analogous to  (\ref{A-c}),
denoting the Pfaffian of an antisymmetric matrix ${\tilde a}$:
  \be\label{a-c}
a^c({\bf z}):=\,\Pf[{\tilde a}]
  \ee
  whose entries are defined, depending on the parity of $N$, in terms
  of a skew symmetric kernel $a(z,w) $ (possibly, a distribution) and a function
  (or distribution) $a(z)$ as  follows:

 For $N=2n$ even
  \be\label{a-alpha-even-n}
{\tilde a}_{ij}=-{\tilde a}_{ji}:=a(z_i,z_j),\quad 1\le i<j \le 2n
  \ee

  For $N=2n-1$ odd
\be\label{A-alpha-odd-n} {\tilde a}_{ij}=-{\tilde a}_{ji}:=
\begin{cases}
a(z_i,z_j) &\mbox{ if }\quad 1\le i<j \le 2n-1 \\
a(z_i) &\mbox{ if }\quad 1\le i < j=2n
 \end{cases}
  \ee
In addition we define $a_0^c =1$.

Integrals $I_1$,$I_2$ and $I_4$ may be considered as analogues of
$\beta=1,2,4$ ensembles \cite{Mehta}. They may be obtained as particular
 cases of $I_3$ as follows:

\noindent
 Integral $I_1(N)$ is a particular case of $I_3(N)$ where in the (A) case
 \be
 a(z_i,z_j)=\sgn(z_i-z_j), \quad a(z)=1
 \ee
  while in case (B)
 \be
a(z_k,z_j)= e^{-\tfrac {\pi i}2 } \sgn(\varphi_k-\varphi_j), \quad
a(z)=e^{-\tfrac {\pi i}4 },
 \ee
 with
$\varphi_i={\text{arg}}\, z_i$.
To prove this we use:
 \bl
 \be
  \Pf\left[\sgn(z_k-z_j)\right]=\sgn\,\Delta^*(z),\quad z_k\in\mathbb{R},
 \ee
 \be
 \Pf\left[\sgn(\varphi_k-\varphi_j)\right]=
 \sgn\,\left(e^{-\tfrac {\pi i}4(N^2-N) }\Delta^*(z)\right),\quad z_k=e^{i\varphi_k}
  \ee
  where $k,j=1,\dots,N$.
 \el

Integral $I_2(N)$ is obtained from $I_3(N)$ by setting
 \be
a(z_i,z_j)=\frac{z_i-z_j}{z_i+z_j}, \quad  a(z)= 1.
 \ee
We use the fact that
  \be
\Delta^*(z) = \Pf\left[\frac{z_i-z_j}{z_i+z_j} \right]
  \ee

Integral $I_4(N)$ is obtained from $I_3(2N)$ as follows. In
case (A) we set
  \be
  \label{delta_kernel}
  a(z_i,z_j)=\frac 12 \left(z_j\frac{\partial}{\partial
  z_j}\delta(z_i-z_j)-(z_i\leftrightarrow z_j)\right)
  \ee
   and in  case (B) we set
  \be
  a(z_i,z_j)=\frac{\partial}{\partial
  \varphi_j}\delta(\varphi_i-\varphi_j).
  \ee

The integrals  containing $\Delta^*$ may be compared with those
defining the partition function of the so-called supersymmetric matrix
integrals, see \cite{Guhr}. Integral $I_2$ defines the partition
function of the so-called ${\hat A}_0$ model (see \cite{Kostov}),
the Coulomb gas model with reflection \cite{LS1},  the 1D Ising model
\cite{LS2}, and correlation functions in the 2D Ising model \cite{Braden}.

To relate these integrals to the 2-BKP hierarchy we introduce
deformations $I_i(N)\to I_i(N;\bt,{\bar\bt})$ through the
following deformation of the measure \be d\nu({ z})\to d\nu({
z}|\bt,{\bar\bt})= b(\bt,\{ z\})b(-{\bar\bt},\{ z^{-1}\})d\nu({z})
\ee where
\begin{equation}
\label{ebb2} b(\bs,\bt)=\exp \sum_{n\ {\rm odd}} \frac{n}{2} s_nt_n
\end{equation}
and
\begin{equation}
\label{bracketz} \{
z\}=(2z,\frac{2z^3}{3},\frac{2z^5}{5},\cdots)\, .
\end{equation}

Below, we show that the generating series obtained by
Poissonization (the grand partition function)
\be
Z_i(\mu \, ;\bt,{\bar\bt})\, =\,b(\bt,{\bar\bt})\sum_{N=0}^\infty
\,I_i(N;\bt,{\bar\bt}) \,\frac{\mu^N}{N!}\,,\quad i=1,2,3,4,
\ee
are particular 2-BKP tau functions (\ref{2-nBKP}).

We also consider the following $2N$-fold integrals:
 \be\label{I5}
I_5(N;\bt^{(1)},\bt^{(2)},{\bar\bt}^{(1)},{\bar\bt}^{(2)}):=\int
\Delta^*_{N}(z)\Delta^*_{N}(y) \prod_{i=1}^{N}d\nu({ z}_i,
y_i|\bt^{(1)},\bt^{(2)},{\bar\bt}^{(1)},{\bar\bt}^{(2)}),
 \ee
where
 \be
  d\nu({
z},y|\bt^{(1)},\bt^{(2)},{\bar\bt}^{(1)},{\bar\bt}^{(2)})=
  \ee
 \[
b(\bt^{(1)},\{ z\})b(-{\bar\bt}^{(1)},\{ z^{-1}\}) b(\bt^{(2)},\{
y\})b(-{\bar\bt}^{(2)},\{ y^{-1}\})d\nu(z,y)
 \]
(here $d\nu(z,y)$ is an arbitrary bi-measure), and show that the
generating series
 \be\label{Z5}
Z_5(\mu\, ;\bt^{(1)},\bt^{(2)},{\bar\bt}^{(1)},{\bar\bt}^{(2)})\,
=\,b(\bt^{(1)},{\bar\bt}^{(1)})b(\bt^{(2)},{\bar\bt}^{(2)})\sum_{N=0}^\infty
\,I_5(N;\bt^{(1)},\bt^{(2)},{\bar\bt}^{(1)},{\bar\bt}^{(2)})
\,\frac{\mu^N}{N!}
 \ee
is a particular case of the two-component 2-BKP tau function
(\ref{2c-2-nBKP}).

 \br
 \label{Z2=Z5}
Note that
\be
Z_2(\mu \, ;\bt,{\bar\bt})
=Z_5(\mu \, ;\bt^{(1)},\bt^{(2)},{\bar\bt}^{(1)},{\bar\bt}^{(2)})
\ee
 if
 \be
 d\nu(z,y) = \delta(z-y)d\nu(z)d\nu(y), \quad
\bt=\bt^{(1)}+\bt^{(2)}, \quad {\bar\bt}={\bar\bt}^{(1)}+{\bar\bt}^{(2)}.
\ee
 \er
The integrals $Z_1(\mu \, ;\bt,{\bar\bt})$, $Z_2(\mu \, ;\bt,{\bar\bt})$,
$Z_4(\mu \, ;\bt,{\bar\bt})$ and
$Z_5(\mu \,  ;\bt^{(1)},\bt^{(2)},{\bar\bt}^{(1)},{\bar\bt}^{(2)})$
 may be obtained as  continuous limits of
$S_1(\bt_\infty,\bt^*)$, $S_2(\bt_\infty,\bt_\infty,\bt^*)$,
$S_4(\bt_\infty,\bt^*)$ and $S_5(\bt_\infty,\bt_\infty,\bt^*)$,
respectively.

\paragraph{Fermionic representation of the integrals.}
To obtain the fermionic representation for the integrals above we
apply (\ref{FermiDelta}).  Expanding the exponentials and applying Wick's theorem
 to each term in the sum gives
 \be\label{Z3F}
Z_3(\mu \, ;\bt,{\bar\bt})=\l 0|\,\Gamma(\bt)\,
e^{\mu^2\int_\gamma\int_\gamma
a(z,y)\phi(z)\phi(y)d\nu(z)d\nu(y)}e^{2\mu \int_\gamma a(z)
\phi(z)\phi_0 d\nu(z)}\, {\bar\Gamma({\bar\bt})}\,|0\r .
 \ee

 Note that
 \be
 Z_3(0\, ;\bt,{\bar\bt})=
 b(\bt,{\bar\bt})=\l 0|\,\Gamma(\bt)\,{\bar\Gamma({\bar\bt})}\,|0\r
 \ee
  was written above in (\ref{ebb2}). In particular we obtain
 \be\label{Z1F}
Z_1(\mu \, ;\bt,{\bar\bt})=\l 0|\,\Gamma(\bt)\,
e^{\mu^2q^2\int_\gamma\int_\gamma
\sgn(\varsigma(z)-\varsigma(y))\phi(z)\phi(y)d\nu(z)d\nu(y)}e^{2\mu
q\int_\gamma \phi(z)\phi_0 d\nu(z)} {\bar\Gamma({\bar\bt})}\,|0\r
 \ee
 where $q=1$ and $\varsigma=\varsigma(z)\,,z\in\gamma$ is  a parameter on $\gamma$, which
 is equal to $\varsigma(z)=z$ when the integration contour is $\mathbb{R}_+$,
 while in the case $z=e^{i\varphi}\in\gamma,\, 0\le \varphi \le
 \theta$, we set $\,\varsigma(z):=\varphi$ and $q=e^{-\frac{\pi i}4}$.

Similarly, we have
 \be\label{Z2F}
Z_2(\mu \, ;\bt,{\bar\bt})=\l 0|\,\Gamma(\bt)\,
e^{\mu^2\int_\gamma\int_\gamma
\frac{z-y}{z+y}\phi(z)\phi(y)d\nu(z)d\nu(y)}e^{2\mu\int_\gamma
\phi(z)\phi_0 d\nu(z)} {\bar\Gamma({\bar\bt})}\,|0\r
 \ee
 As a specialization of (\ref{Z3F}), by choosing $a(z,w)$ as in (\ref{delta_kernel})
 and $a(z)=0$, we obtain
  \be\label{Z4F}
Z_4(\mu \, \bt,{\bar\bt})=\l 0|\,\Gamma(\bt)\, e^{4\mu\int_\gamma z
\frac{\partial\phi(z)}{\partial z}\phi(z)d\nu(z)}\,
{\bar\Gamma({\bar\bt})}\,|0\r .
 \ee

Finally, in terms of two-component fermions we have
 \bea
 \label{Z5F}
&&
Z_5(\mu;\bt^{(1)},\bt^{(2)},{\bar\bt}^{(1)},{\bar\bt}^{(2)}
 \cr
&=&\frac 1c\l
0|\,\Gamma^{(1)}(\bt^{(1)})\Gamma^{(2)}(\bt^{(2)})\,
e^{2\mu\int\int\,\phi^{(1)}(z)\phi^{(2)}(y)d\nu(z,y)}\,
{\bar\Gamma}^{(1)}({\bar\bt}^{(1)}){\bar\Gamma}^{(2)}({\bar\bt}^{(2)})\,|0\r
,
 \eea
where
$c:=Z_5(0 \, ;\bt^{(1)},\bt^{(2)},{\bar\bt}^{(1)},{\bar\bt}^{(2)})$ is
the normalization factor.

Formulae (\ref{Z3F}),(\ref{Z1F}),(\ref{Z4F}), and (\ref{Z5F})
should be compared with (\ref{S3F}),(\ref{S1F}),(\ref{S4F}), and
(\ref{S5F}), respectively.The fermionic representations of integrals $Z_1$ and $Z_4$
are similar to the results of \cite{Johan99} for ensembles of real
symmetric and  self-dual quaternionic random matrices.

%%%%%%%%%%%%%%%% Section 4  Discussion %%%%%%%%%%%%%%%%

\section{Discussion}

We have presented five types of multiple sums and multiple
integrals which are related to  tau functions of BKP type
hierarchies (BKP and two-component BKP tau functions for multiple
sums, 2-BKP and two-component 2-BKP tau functions for multiple
integrals). Certain of these sums and integrals have known applications in
mathematics and physics; we believe that all of them may prove to be
of use in various probabilistic models.  The techniques of free fermion calculus
and  integrable systems may be applied to study various properties of these
sums and integrals. Multicomponent BKP
tau functions \cite{KvdLbispec} may be further used to construct
models of Pfaffian processes (cf. \cite{paper5}). The
results of this work should be be compared with analogous results for
the so-called charged BKP case (see \cite{KvdLbispec}) studied
in \cite{Johan99} and \cite{sQQ}.

\section{Appendix. An application of the integral $Z_2(N)$
 (Harry Braden).}

Some of the considerations of this paper were motivated by the
following observation of Harry Braden (\cite{Braden}).

``The application of the integrals $Z_2$ I have in mind deals
with an Ising model correlation function when there is a thermal
perturbation from critical temperature. In the scaling limit this
is described by a system of Majorana fermions and makes connection
with the fermionic representation under consideration. McCoy,
Tracy and Wu showed this limit was described a massive ($m=T-T_c$)
field theory whose correlations were governed by a radial
Sinh-Gordon equation that under a change of parameters is
$P_{III}$. Another approach to the same correlator is via form
factors. This approach yields the correlator in terms of an
infinite sum of multiple integrals. For the case at hand this
gives a Euclidean correlation function like
\begin{align*}
G(r):=<\mathcal{O}(x)\mathcal{O}(0)>&=\sum_{n=0}\sp\infty \int
\frac{\prod_{i=1}\sp{n} d\beta_i}{n! (2\pi)\sp{n}}
<0|\mathcal{O}(x)|\beta_1\ldots\beta_n><\beta_n\ldots\beta_1|\mathcal{O}(0)|0>\\
&=\sum_{n=0}\sp\infty \int \frac{\prod_{i=1}\sp{n} d\beta_i}{n!
(2\pi)\sp{n}}|F_n(\beta_1\ldots\beta_n)|\sp2
\exp(-mr\sum_{i=1}\sp{n}\cosh\beta_i).
\end{align*}
Here $\beta_i$ are rapidities and $x=(x_0,x_1)$,
$r=\sqrt{x_0^2+x_1^2}$. If we use the minimal form factor
$$F_n\sp{\mathrm{min}}(\beta_1\ldots\beta_n)=\prod_{i<j}\tanh\left(\frac{\beta_i-\beta_j}{2}\right)$$
and appropriate normalizations we get
\begin{equation*}
G\left(\frac{r}{m}\right)=\sum_{n=0}\sp\infty\frac1{n!}\frac{1}{(2\pi)\sp{n}}
\int_0\sp\infty\prod_{i=1}\sp{n}\left( \frac{ dx_i}{x_i}\,
e\sp{-r(x_i+1/x_i)}\right)\prod_{i<j}\left(\frac{x_i-x_j}{x_i+x_j}\right)\sp2.
\end{equation*}
As I havent been specific about the precise correlator, we also
are interested in
\begin{equation}
G_\pm\left(\frac{r}{m}\right)=\sum_{n=0}\sp\infty\frac1{n!}\frac{(\pm1)\sp{n}}{(2\pi)\sp{n}}
\int_0\sp\infty\prod_{i=1}\sp{n}\left( \frac{ dx_i}{x_i}\,
e\sp{-r(x_i+1/x_i)}\right)\prod_{i<j}\left(\frac{x_i-x_j}{x_i+x_j}\right)\sp2.
\label{gpm}
\end{equation}
The identity
$$\det\left(\frac1{x_i+x_j}\right)=\frac1{2\sp{n}\,x_1\ldots x_n}
\prod_{i<j}\left(\frac{x_i-x_j}{x_i+x_j}\right)\sp2$$ provides a
connection with a Fredholm determinant. I have looked at
expansions of (\ref{gpm}). The difficulty is getting a convergent
expansion. The first term is in terms of $K_0(r)$. Such an
expansion will provide one for (a particular) solution of the
Painlev\'e equation."

\section*{Acknowledgements}
The authors are grateful to T. Shiota and J. J. C. Nimmo for
useful discussions. One of the authors (A.O.) thanks A. Odjievicz
for kind hospitality during his stay in Bialystok in June 2005.
Both A.O and J.vdL. thank the CRM, Montr\'eal, Canada for the kind
hospitality during their stay in January 2006, where the main part
of this paper was written.

 \end{document}